\documentclass[journal]{IEEEtran}

\ifCLASSINFOpdf
\usepackage[pdftex]{graphicx}

\else

\fi

\usepackage{float}
\usepackage[cmex10]{amsmath}
\usepackage{amssymb}   
\usepackage{amsxtra}
\usepackage{amscd}
\usepackage{amsthm}
 \usepackage{amsxtra}
 \usepackage{amscd}
 \usepackage{steinmetz}
 \usepackage{amsthm}
\usepackage{tikz}
\usetikzlibrary{matrix}
\usepackage{stfloats}
\usepackage{graphicx, subfigure}   
\usepackage{relsize}
\usepackage{algorithm,algpseudocode}

\usepackage{array}  
\usepackage{authblk}
\usepackage{multirow}  
\usepackage{arydshln}

\usepackage{color}

\usepackage{cite}
\renewcommand{\thefootnote}{\arabic{footnote}}

\usepackage{balance}

  \author{
  Emre Arslan,~\IEEEmembership{Graduate Student Member,~IEEE,}
  Ali Tugberk Dogukan,~\IEEEmembership{Graduate Student Member,~IEEE,}
  Fatih Kilinc,~\IEEEmembership{Graduate Student Member,~IEEE,}
 Ahmet Faruk Coskun,~\IEEEmembership{Graduate Member,~IEEE,} \newline and
Ertugrul Basar,~\IEEEmembership{Fellow,~IEEE}
\vspace{-0.6cm}  
  
  \thanks{Emre Arslan and Ahmet Faruk Coskun are with 6GEN. Lab of
Turkcell Iletisim Hizmetleri Inc., Maltepe 34854, Istanbul, Turkiye.  Ali Tugberk Dogukan and Fatih
Kilinc are with ULAK Communications,
Pendik 34906, Istanbul, Turkiye. Ertugrul Basar is with
Communications Research and Innovation Laboratory (CoreLab), Department
of Electrical and Electronics Engineering, Koc¸ University, Sariyer 34450,
Istanbul, Turkiye. Emre Arslan, Fatih Kilinc and Ali Tugberk Dogukan are also members of CoreLab. 
\mbox{Email: \{emre.arslan@turkcell.com.tr}, {tugberk.dogukan@ulakhaberlesme.com.tr}, {fatih.kilinc@ulakhaberlesme.com.tr}, {coskun.ahmet@turkcell.com.tr}, {earslan18@ku.edu.tr}, {and ebasar@ku.edu.tr}\}
  } 
  
}

\begin{document}

	\title{{Reconfigurable Intelligent Surface Identification in Mobile Networks: Opportunities and Challenges}}

	\maketitle	
	
	\begin{abstract}

The advent of the sixth generation (6G) wireless networks heralds a transformative era for mobile communication, where the integration of cutting-edge technologies like Reconfigurable Intelligent Surfaces (RISs) is paramount in addressing the burgeoning demands for energy efficiency, high data rates, reliable connectivity, and enhanced coverage in densely populated areas. RISs have attracted the attention of academia as well as the industry and emerged as a beacon of innovation, offering a novel paradigm to reconfigure the wireless propagation environment beneficially, thereby enhancing the overall network performance. It is envisioned that the deployment of numerous RISs in a mobile network will serve user equipments (UEs) to boost various key performance indicators (KPIs), with
lower energy consumption compared to alternative solutions
(e.g., relays, and network-controlled repeaters). However, the knowledge of whether or not a UE is being served through an RIS and if so, which RIS it is being served by is crucial and beneficial for various network planning and operational reasons. In this paper, we address the importance and the benefits of RIS identification in a mobile network. Additionally, we guide the readers and researchers by introducing alternative methods to enable RIS identification. Through the lens of this research, we unveil and shed light to various benefits, the challenges, and future opportunities in the identification of RISs serving mobile users in complex network environments, highlighting the necessity for advanced identification strategies to fully harness the potential of RIS technology in next-generation wireless systems.

	\end{abstract}


	\IEEEpeerreviewmaketitle
		\IEEEpubidadjcol
		
	\renewcommand{\thefootnote}{\fnsymbol{footnote}}

	\section{Introduction}
 	

\IEEEPARstart{A}{dvancing} into the future, sixth-generation (6G) wireless networks are poised to revolutionize communication technologies with their promise of terabit-per-second speeds, enhanced privacy, virtually zero latency and ultra-reliability even in densely populated areas. 
Numerous unique technologies and their coexistence such as novel waveform designs, multiple access techniques, reflective metasurfaces, machine learning (ML), and much more are investigated and developed to realize these 6G goals. One critical technology for future wireless networks is the integration and deployment of reconfigurable intelligent surfaces (RISs) \cite{8910627}. Gaining popularity for their ability to control and optimize the wireless propagation environment, RISs consist of programmable elements that can dynamically alter the phase, amplitude, and polarization of the incoming electromagnetic waves \cite{6GRIS}. 
RIS technology offers substantial benefits in terms of network coverage, signal strength, and data secrecy, improving the overall user experience (UX), as well as system capacity and energy efficiency. 
On the other hand, there are
many practical concerns awaiting solutions regarding the installation,
integration, and management of RIS equipment \cite{WPMC}. Among these,
the main challenges might be listed as
\begin{itemize}
\item Clarifying practical deployment benefits and drawbacks compared to previous network equipment like NCRs and RF repeaters,\cite{repeater},
\item Ensuring effective coordination during operation (control by base stations (BS, RIS itself, or user equipment (UE)), 
\item Developing optimization algorithms for RIS configuration with or without cascaded channel state information (CSI).

\end{itemize}

Despite these hurdles, RIS stands as a cornerstone technology for 6G, promising to transform wireless communications by enabling smarter and more efficient network behaviors \cite{ETSI, arxivbasar}.

RIS is envisioned for extensive deployment by operators in various environments, from urban centers to rural areas, enabling networks to overcome physical obstructions and distance limitations. RIS will be useful in many indoor/outdoor applications, such as optimizing signal distribution in large public venues and enhancing communication reliability in emergency response scenarios. This strategic deployment enhances 6G network capacity and efficiency, paving the way for innovative services leveraging robust connectivity \cite{deployment}. Additionally, RIS serves as an anchor for ISAC applications \cite{ISAC_RIS}, enhancing their diversity and accuracy, making RIS a versatile asset for 6G technology.


With the increasing deployment of RIS, identifying which RIS a UE interacts with is critical \cite{Aymen}. RIS identification, determining which RIS is affecting a user's signal, is vital for optimizing real-time communication, enhancing signal quality, network coverage, capacity, security, and localization accuracy. It also plays a crucial role in network planning, interference management, handover, and roaming. Accurate RIS identification can significantly improve KPIs such as throughput, security, reliability, energy efficiency, and localization in 6G networks. However, this requires sophisticated detection and communication protocols to ensure accurate RIS recognition, especially in environments with multiple RISs.

Despite its importance, the domain of RIS identification remains mainly underexplored. Most existing research focuses predominantly on the design and optimization of RISs without adequately addressing the challenges associated with their discovery and (mis)identification in an operational network. Identifying an RIS involves determining not just its presence but also its specific influence on the signal path, a task complicated by the passive nature of most RIS implementations and the subtle variations in the signal characteristics they induce.
As of now, the identification of RISs in wireless networks remains largely underdeveloped, primarily due to their passive nature. Unlike active network components such as BSs and repeaters, which identify themselves through broadcasts like SSIDs, RISs do not inherently transmit any identification signals. Current approaches proposed for active network components
might leverage signal strength measurements, timing
advances, or complex CSI, which can be cumbersome and
imprecise, particularly in dynamic and cluttered environments. This challenge introduced by the passive nature of RIS necessitates innovative approaches to it's identification, borrowing concepts from technologies like RF identification (RFID) and near-field communication (NFC), which rely on distinct signals to recognize tags or devices.


To adapt these strategies for RISs, potential methods include embedding unique signatures within the phase shifts or amplitude variations imparted on reflected signals, detectable by receiving devices to identify the specific RIS. Alternatively, ML algorithms could analyze and classify subtle modifications made by different RISs based on extensive training with known configurations. Developing these identification techniques is crucial, requiring solutions that integrate effectively with the dynamic nature of 6G networks while maintaining scalability and efficiency.


For RIS identification, approaches can be broadly categorized based on whether they are implemented on the BS side or directly over-the-air (OTA) by the RIS. On the BS side, techniques may focus on marking signal patterns transmitted over the RIS prior to transmission to indicate which surface is in use. This method offers centralized control, which can simplify management and optimization but may struggle with scalability in environments dense with multiple RIS. Alternatively, OTA methods involve receivers detecting unique modifications in the signal, such as subtle changes in phase or amplitude, that indicate specific RIS engagement. These techniques allow for more distributed processing and can scale effectively as network density increases, although they place higher demands on receiver sensitivity and complexity. Both approaches aim to balance the identification accuracy with network performance, each offering distinct advantages and challenges that need consideration in diverse operational contexts.




Inspired by the potential and popularity of RIS-empowered systems resonated in academia and industry to realize smart wireless environments, this study provides a unified RIS identification framework and a thorough exploration of the identification process, presenting its fundamental principles and explaining the substantial advantages it offers to wireless communication networks. Our comprehensive discussion aims to clarify the operational impact and strategic importance of accurate RIS identification. We introduce three promising research directions and preliminary techniques for realizing RIS identification in wireless networks in different domains each with various issues and considerations. Our discussion is extended with a critical examination of the prevailing issues and challenges inherent in RIS identification. Furthermore, we propose potential strategies and future research avenues, which are crucial for providing a deeper and insightful perspective on the practical
evaluation of RIS-assisted wireless networks.



  
     
     
        

The remainder of this article is organized as follows. Section II discusses RIS identification in a mobile network and the benefits it brings. Section III then introduces some preliminary RIS identification techniques considering different domains to provide insight and trigger future alternative solutions. Section IV highlights and outlines some of the research challenges which may need to also be considered when performing RIS identification. Finally, the article is concluded in Section V.

\begin{figure*}[ht]
	
		\centering
		\includegraphics[width=.85\textwidth,height=8.4cm,scale=.1]{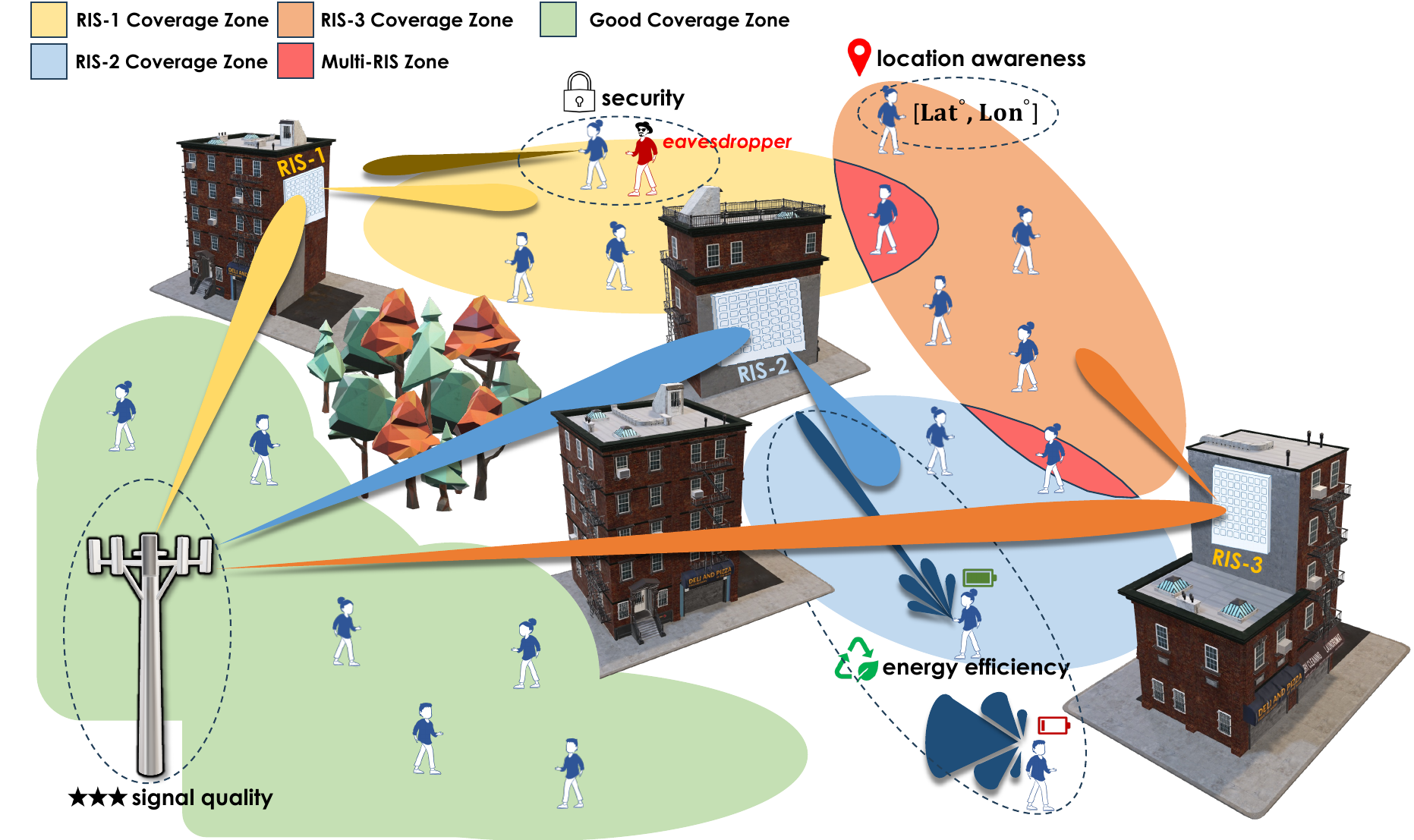} 
		\caption{Representative sketch for a densely
populated cellular environment utilizing multiple RISs}
		\label{fig:Fig.1}
\end{figure*}

\section{ RIS Identification}

In the intricate urban tapestry where skyscrapers and concrete establishments disrupt direct communication pathways, RIS-empowered communication emerges as an enabling technology for enhanced connectivity. The urban scenario depicted in the Fig. 1 is representative of modern cities where dead zones with poor signal quality are commonplace. In such zones, the direct line-of-sight (LoS) to the BS is often obstructed, leading to deteriorated communication performance, especially for higher frequency
bands established by 5G (e.g., FR2) and beyond (e.g., FR3
and D-band). Deploying RISs within these areas is a strategic move to mitigate such connectivity challenges, with each RIS assigned a unique identification code, enabling not just signal reflection and refraction but also intelligent environment-aware signal enhancement.

RIS identification becomes paramount when a user's device enters such a zone and should first detect if it is being served by an RIS or not. Beyond mere detection, the device must ascertain the specific RIS from which it is receiving the signal. This identification process is crucial as it provides many benefits with the identification information and determines the quality of service the UEs experiences, which will be discussed in the following subsections.
However, there could be scenarios where the UE is in the coverage of multiple RISs so called multi-RIS zones.
These are regions where the coverage zones of multiple RISs overlap. In these areas, the signals from different RISs might intermingle, similar to inter-cell interference, leading to a scenario where the UE device may struggle to discern which RIS it is interacting with. This confusion may cause the device to either oscillate between RISs or, worse, connect to an RIS with a weaker signal or poorer performance relative to its position. Hence, RIS identification can aid in detecting whether a UE is served by a RIS and also in distinguishing which RIS it is served by when in a multi-RIS zone.

RIS identification thus entails an approach where the device must employ algorithms capable of distinguishing signals and RIS IDs based on their unique BS or RIS-induced modifications. This may involve detecting subtle differences in signal phase, power, polarization, or timing introduced by the BS or RIS's reflective properties. For 5G and beyond, the integration of RISs is anticipated to adhere to stringent standards that ensure seamless operability, optimally-mitigated interference, and maximal efficacy within the network. The RIS identification process is thereby not only a question of signal discernment but also one of compliance with the protocols and performance benchmarks set forth by these advanced communication standards. 

RIS identification techniques may assist in detecting multiple RISs and in such a scenario choosing the RIS that provides optimal performance.
In the context of 5G networks, by identifying the correct RIS, a UE can optimize its connection to exploit a wider range of 5G's capabilities, such as high data rates, higher efficiency, improved reliability, and reduced latency by directing its beam towards the dedicated RIS.

In dense urban environments, RIS technology enables networks to proactively address the diverse demands of modern wireless communication. The convergence of RIS with 5G standards represents the future, where devices actively participate in a coordinated flow of data, guided by RISs. Detailed information on beneficial use cases for RIS identification in beyond 5G wireless communications is provided in the following subsections.


\subsection{Signal Quality Enhancement}

RIS identification has a direct impact on enhancing signal quality by allowing precise control and optimization of the communication path between the UE and BS. Specifically, when a UE can identify the RIS it is connected to, it can engage in more effective uplink communication by directing its transmissions towards the RIS, thus improving the connection quality and efficiency with the BS. This targeted approach reduces signal degradation and optimizes bandwidth usage, crucial for high-demand applications. Additionally, UEs can fine-tune their device settings, such as transmission power and beam selection/direction, based on the characteristics of the identified RIS, addressing factors like signal strength and environmental interference. 

Moreover, RIS identification may aid in network-wide signal quality improvement by informing the BS about which RISs are actively serving UEs. This information allows the network to prioritize resources and adjust configurations, ensuring that users connected to a particular RIS receive the best possible signal. In congested network environments, this can mean dynamically redistributing network load or enhancing signal fidelity for users in high-density areas, leading to a more balanced and effective network performance.

\subsection{Optimized Security and Secrecy}

In the wireless communication field, RISs have revolutionized not only signal optimization but also communication security and secrecy \cite{security}. Identifying the specific RIS serving a UE can enhance security in wireless networks, allowing for dynamic adaptation to threats and enabling secure operations for sensitive services like e-commerce and data exchange. For example, in high-security zones, devices can activate enhanced security measures, such as advanced encryption and authentication, based on their RIS connectivity, thus fortifying the network against unauthorized access and eavesdropping. Furthermore, UEs may direct signals to the identified RIS to reduce interception risks, safeguarding the legitimate data path.
Moreover, the unique ability of RIS to manipulate signal phases and amplitudes adds a layer of physical layer security, concealing communications from eavesdroppers and representing a significant step toward proactive security measures in wireless networks \cite{secrecy1,secrecy2}.

\subsection{Energy Efficiency}

The integration of RIS in mobile networks could significantly enhance energy efficiency by enabling precise RIS identification and interaction. As illustrated in Fig. 1, when the UE identifies which RIS it is served through the UE can perform directed beamforming hence, focusing energy use on necessary paths toward the RIS rather than less efficient non-line-of-sight (NLOS) or omnidirectional transmissions, thus reducing power wastage. Devices can adjust their power output based on the RIS coverage, leading to battery conservation and sustained service quality, especially crucial in mobile environments. RIS identification also aids in network-wide energy management, aiding operators to determine the frequently used RISs,  allocate resources intelligently and activate RIS panels as needed hence, enhancing network efficiency and contributing to environmental sustainability. 

\subsection{Localization and Sensing
}


One of the most prominent benefits of RIS identification complementary to the 6G vision is the enhancement of localization in wireless networks by enabling UEs to determine their rough location based on the specific RIS connected to. This process allows for a granular level of location awareness, beneficial in environments where traditional GPS struggles, such as densely built-up urban areas or large and complex indoor spaces like malls and airports. 

For example, by identifying the RIS that assists the nearby region, a UE can ascertain that it is within a particular zone of a shopping center, enabling tailored navigation services and location-specific information such as store promotions or directory assistance. This identification can also streamline emergency responses; when a user can pinpoint and communicate their location via the connected RIS, emergency services can rapidly locate and reach the individual, reducing response times and potentially saving lives in case of disasters and possible casualties. As illustrated in Fig. 1, when the user served by RIS 3 identifies the RIS, it can obtain rough information that it is in the orange coverage area and have location awareness. 
Additionally, from a time difference of arrival localization perspective, identifying all surrounding RISs also might be utilized for triangulation purposes in case of coordinated RIS deployments. 
This level of service personalization not only improves user engagement but also opens avenues for innovative service offerings in various sectors, making RIS identification a cornerstone of next-generation wireless communication systems that aim to be more immersive, efficient, and user-focused.

\subsection{Smarter Interference Management}

Intelligent interference management, enabled by accurate RIS identification, allows users to actively participate in optimizing network performance. By recognizing the serving RIS, devices can dynamically adjust their transmission parameters, such as power levels and beam directions, to minimize inter-cell interference. This real-time adaptation is crucial in environments with multiple RISs, where the potential for signal conflict is high. Targeted communication towards the identified RIS reduces unnecessary signal spread, avoiding interference with other users in the vicinity. This approach not only improves the individual user experience by ensuring more reliable connections but also enhances the overall network efficiency by reducing cross-interference among users connected to adjacent RISs.

\section{RIS Identification Techniques}

This section briefly introduces preliminary and pioneering
identification approaches for RISs in wireless networks
in successive subsections. The following proposed identification methods are briefly summarized in Table I.

\begin{table*}
    \caption{Comparison of Proposed Identification Techniques}
    \label{tab:my_label}
    \centering
    \begin{tabular}{c c c c c }
    \hline    
        {\bf{Method}}   & {\bf{Domain}}     & {\bf{Computation Complexity}} & {\bf{Precision/Robustness}} & {\bf{Waveform}}  \\ \hline \hline
        
        {\bf{Amplitude Modulation Sequence}}      & Time & Low        & Low  & Single-/Multi-carrier  \\
        
        {\bf{Spectral Fingerprinting}}   & Frequency & Moderate       & Moderate & Multi-carrier       \\
        
        {\bf{BS Side Watermarking}} & Code and Spatial       & High       & High  & Single-/Multi-carrier      \\
            \hline 
    \end{tabular}
\end{table*}

\subsection{Amplitude Modulation Sequence}


The principle behind this technique involves the alteration of the impinging signal's amplitude by the RIS to embed a recognizable pattern. Specifically, the RIS can selectively boost or attenuate the signal's strength by either coherently and destructively manipulating the phases of its elements or switching off a subset of its elements at predetermined intervals, thereby reducing the signal reflection or redirection during those periods. This creates a signature in the signal's amplitude that can be detected and decoded by the receiver. This can be applied to both single-carrier and multi-carrier systems.




Fig. 2 illustrates this concept by depicting multiple RIS units, each modulating the signal's amplitude in a distinct manner, creating identifiable power variations at UEs' receivers. These variations are encoded into the signal as unique identifiers, enabling each UE to discern and authenticate the RIS in communication, thereby ensuring a secure and efficient network operation.

\begin{figure}[t]
		\includegraphics[width=0.95 \columnwidth,height=5.2cm]{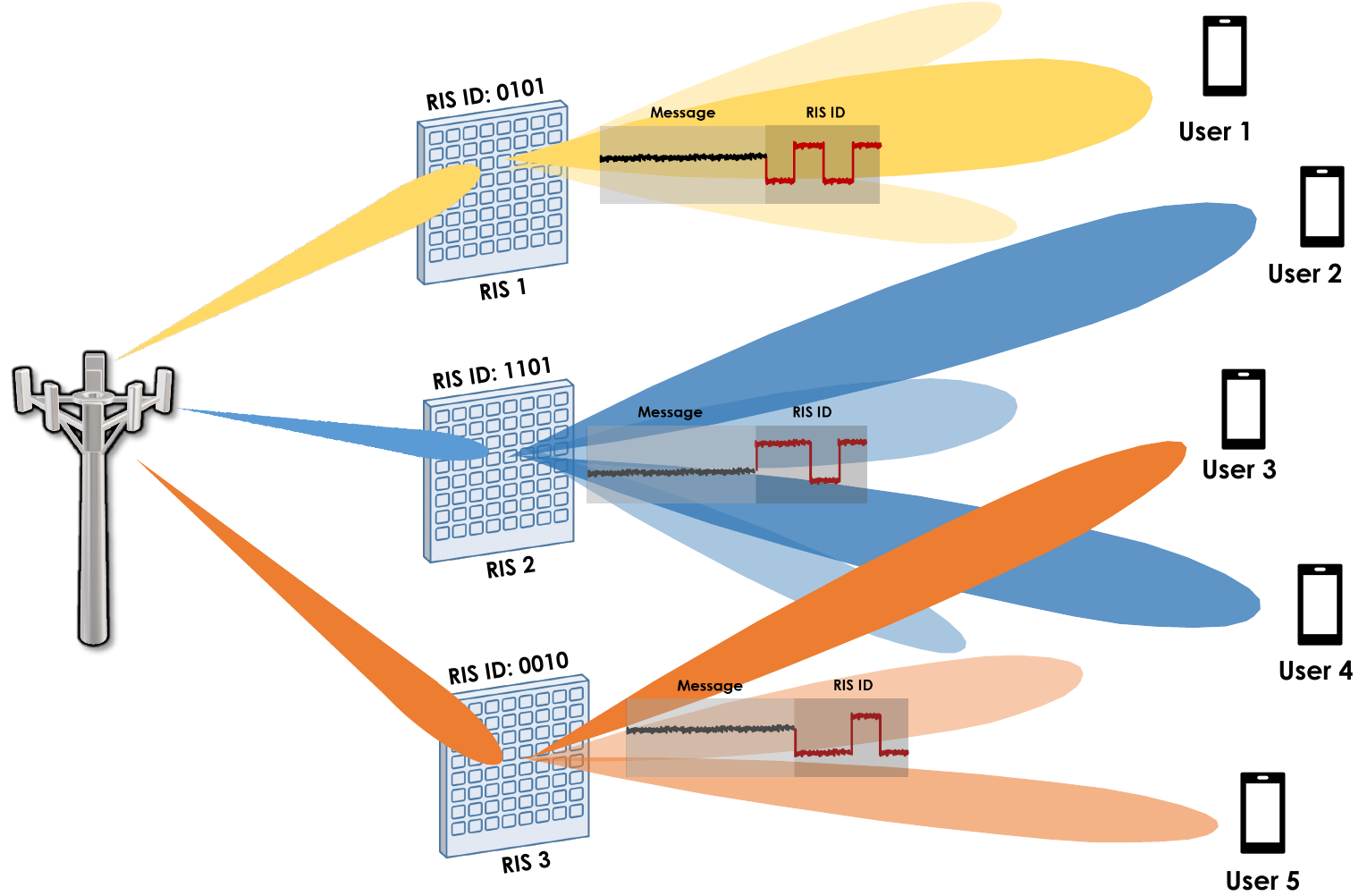} 
		\caption{Illustration of amplitude modulation sequence technique for RIS identification.}
		\label{fig:Fig1}
\end{figure}

\subsection{Spectral Fingerprinting}

The spectral fingerprinting method provides a refined approach that leverages the frequency domain characteristics of a multi-carrier signal.   The central premise of this method is the strategic enhancement of specific subcarriers or subsets by a specific RIS for identification. The RISs can adjust themselves according to the channels of certain subcarriers and boost them. By doing so, each RIS imprints a distinct spectral signature on the signal that can be easily recognized and distinguished at the receiver end with subcarriers' orthogonality feature. This is depicted in Fig. 3, where the power distribution across subcarriers is noticeably altered at each UE after the
reception of all existing RIS-assisted signals.

In the scenario sketched in
Fig. 3, based on the the instant location of the UEs and the existing
RISs, the frequency spectrum given for each UE
includes information about the existing RISs’ dominance. The
peaks corresponding to the intensity of RIS signatures help
determine and manage the supporting RISs for each UE.
The presence of distinct peaks correlates to the respective RISs active during transmission. Through careful analysis of these power distributions, the UE can discern which RISs have been engaged and, by extension, infer their relative efficacy and proximity. This RIS identification
approach might momentarily face misidentifications
due to spectral notches that can occur in frequency-selective
propagation environments, hence parameters such as number of symbols to consider for the identification process, bandwidth and size of subcarrier grouping for spectral fingerprinting may be considered to increase robustness of the system.

\begin{figure}[t]
		\includegraphics[width=1 \columnwidth,height=5cm]{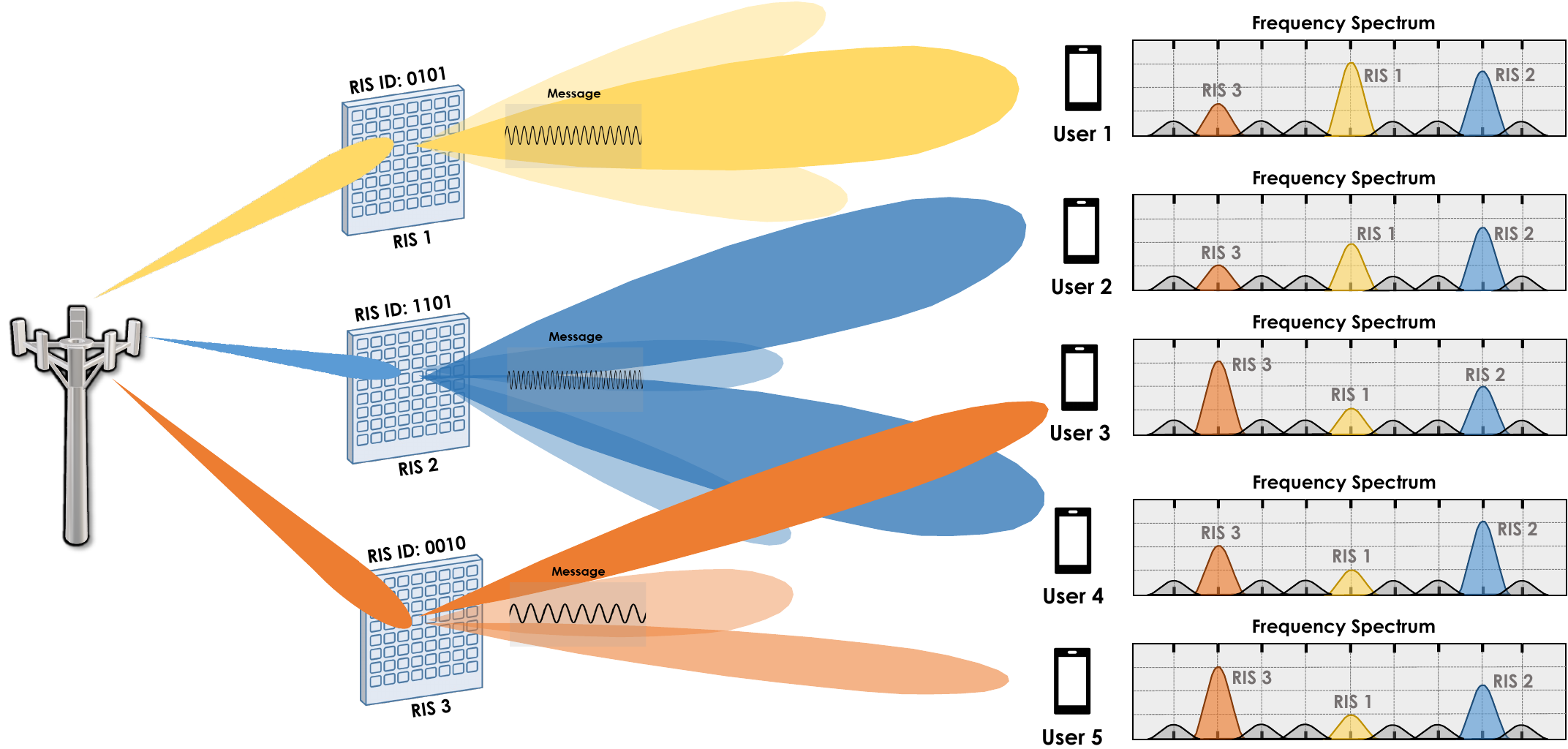} 
		\caption{Illustration of spectral fingerprinting technique for RIS identification}
		\label{fig:sss}
\end{figure}

\subsection{BS Watermarking
}

This core concept revolves around the BS embedding a unique identification sequence within the single or multi-carrier signals directed towards the various RISs, as depicted in the Fig. 4. The BS crafts these signals, \( x_1, x_2, \) and \( x_3 \), with distinct characteristics, akin to a digital watermark. This could be a set of orthogonal codes, cyclic shifts, or specific phase shifts, each serving as a beacon for the UE to recognize the signal's origin and path. Such distinct modifications are not merely arbitrary; they are meticulously designed 
sequences with exceptional auto-correlation and poor cross-correlation properties, ensuring minimal interference and optimal path discernment by the UE. When the BS broadcasts these signals, the UE's task is to correlate the received sequences and pinpoint the exact RIS through which the signal was reflected. This is similar to a detective matching fingerprints at a crime scene, each unique sequence leads back to a specific RIS, allowing the UE to 'localize' its communication pathway.



Fig. 4 illustrates this concept where the BS is the signal origin, with each RIS reflecting these signals to the UE. The unique identification sequences encoded within the signals are the critical factor enabling each UE to discern and trace the signal path back to its RIS of origin. The type and length of the sequences are important when designing such a system. This not only enhances the accuracy of localization systems but also adds a layer of security and efficiency in network communication, as each RIS can be distinctly and accurately identified by its 'watermarked' signal.

In Fig. 5 we provide computer simulation results to demonstrate the effects the parameters in the system has on the identification performance. For this simulation, there are three RISs with varying number of elements $N$ where the BS communicates with the UE through only the RISs under a single carrier system. Considering a coordinate plane the BS, RIS 1, RIS 2, RIS 3, and UE are located at (0,0), (15,0), (0,20), (20,17), (8,17), respectively. Using Gold sequences with varying lengths and QPSK modulation, we assume that there is Rayleigh fading channel and the UE receives the signal from all RISs while RIS 2 is the primary serving RIS. It can be clearly observed that as the code length and number of RIS elements ($N$) increases, the misidentification rate decreases hence, a better identification performance. 


\begin{figure}[t]
		\includegraphics[width=0.95 \columnwidth,height=5cm]{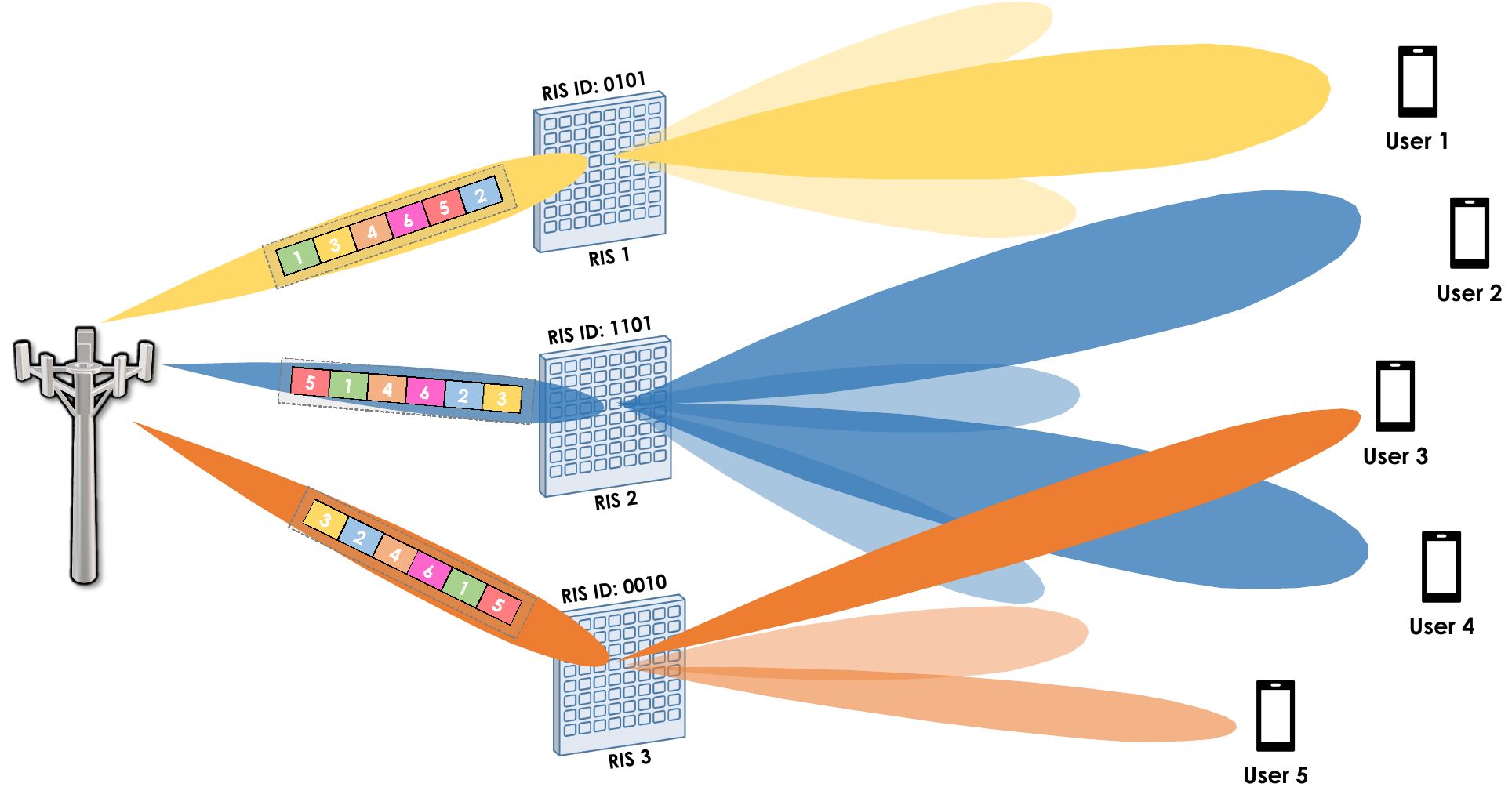} 
		\caption{Illustration of BS watermarking technique for RIS identification.}
		\label{fig:Fig1}
\end{figure}

\subsection{Further possible methods}

Further RIS identification methods can use the effects of element spacing in RIS hardware. The angular width of the main lobe, grating lobes, and sidelobe levels can create unique patterns detectable at the receiver. The frequency response, influenced by element spacing, may also show distinct patterns across frequencies. Additionally, frequency offsets or harmonics from temporal modulation can serve as unique watermarks for secure communications. Implementing these methods requires extensive data collection and complex algorithms but offers promising avenues for precise RIS identification. However, these attributes present implementation challenges compared to previous methods. Obtaining these attributes requires comprehensive data collection and a wide range of equipment. Utilizing them necessitates cooperation between UEs and more complex identification algorithms.

\begin{figure}[t!]
		\includegraphics[width=0.95 \columnwidth,height=6.2cm]{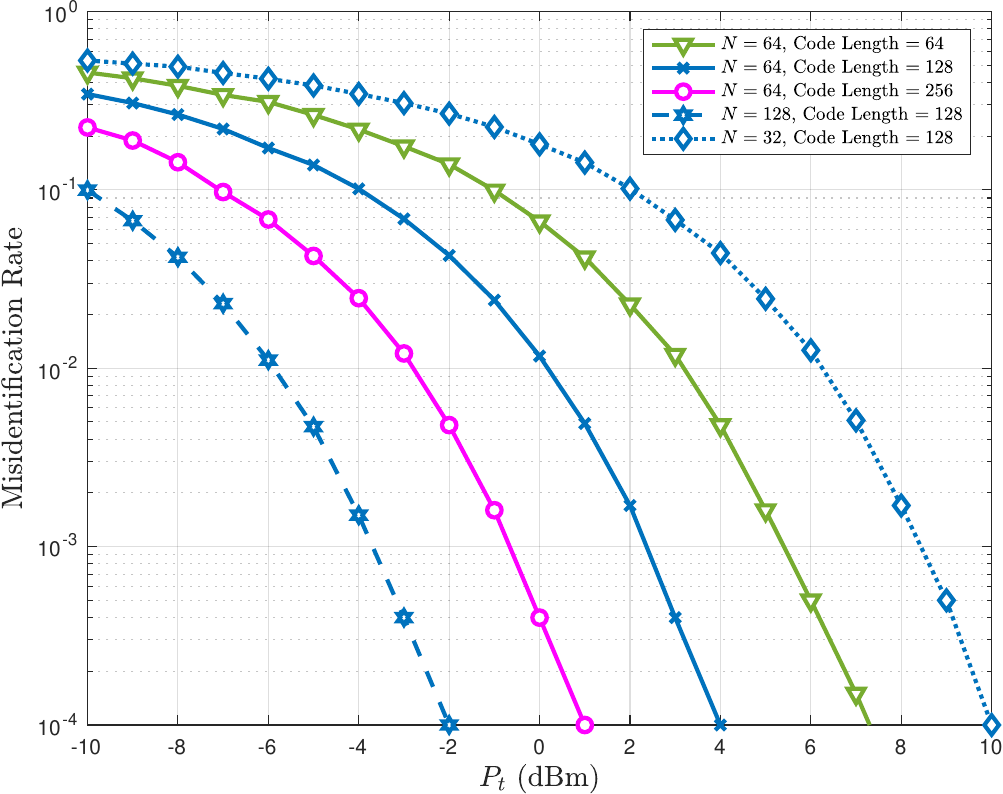} 
		\caption{Performance of BS watermarking technique for varying parameters.}
		\label{fig:Fig1}
\end{figure} 

        
        
        



\section{Future Directions and Open Challenges}


As wireless networks evolve, the expected density and scalability of RIS deployment are expected to significantly increase \cite{2023RISMarket}. Managing and identifying a large number of RIS panels, especially in dense urban environments or large indoor spaces, poses a substantial challenge. Ensuring that each RIS can be uniquely identified without creating an overly complex or resource-intensive process is crucial. The scalability challenge also extends to the dynamic nature of wireless environments, where the configuration of RIS panels may need to change frequently to adapt to varying network conditions and user demands. It is imperative to address the mechanics of the identification process, its timing, and the strategic implementation to ensure seamless network integration and optimal performance. The following subsections introduce
some prominent future directions and open challenges
regarding the identification of RISs.

\subsection{Identification Period}

Determining an optimal RIS identification period presents technical challenges and opportunities for innovation in wireless networks. The identification period, or time allocated for a device to identify the RIS, is crucial for supporting various operational requirements and balancing identification reliability with overhead. The decision to implement a distinct identification period requires careful consideration of its benefits and drawbacks, as well as exploring alternatives for seamless integration without sacrificing performance. 

RIS identification can be done with a dedicated period or simultaneously with data transmission. A dedicated period allows for focused assessment of signal characteristics and the RIS's impact, leading to better decisions on beamforming, power allocation, and adaptive mechanisms. However, it consumes valuable time that could be used for data transmission, especially during high-demand scenarios, potentially increasing latency and reducing throughput. The dynamic nature of mobile environments means the optimal identification period may vary, requiring adaptive strategies that complicate network management.

\subsection{Misidentification}



When a device erroneously identifies a non-serving RIS as its source of signal enhancement, the primary concern is the potential for reduced user and network performance. This is particularly problematic in border areas between RIS zones called multi-RIS zones, where signals from adjacent RISs might overlap. In such cases, a device might receive conflicting signal enhancements, leading to a degradation in signal quality and overall communication reliability. Moreover, false RIS detection can result in suboptimal resource allocation, where network resources are wasted on attempting to optimize a connection with an incorrect RIS, thereby reducing the network's overall efficiency. Some strategies such as enhanced signal processing techniques to distinguish reflected signals, strategically planning the deployment of RISs in a network to mitigate the risk of false detection and/or misidentification, ML-based solutions, and user feedback mechanisms may aid and are future directions in the misidentification problem.

\subsection{Optimal Identification Domain and Technique}



The choice of domain for RIS identification (time, power, frequency, or spatial) significantly influences the system's efficiency and reliability. Time-domain identification, for instance, may involve specific time slots dedicated to transmitting RIS IDs, potentially impacting data throughput due to the allocation of resources for identification purposes. Power-domain strategies could involve varying the power levels to encode RIS IDs, posing challenges in dynamic environments where signal strength can fluctuate due to mobility or obstacles. Frequency-domain identification offers the advantage of embedding IDs within certain frequency bands, though this requires careful spectrum management to avoid interference with data transmission. Spatial domain identification, leveraging the unique spatial signatures that RIS can impart on signals, promises a less intrusive method, though it demands sophisticated signal processing techniques.

As we advance, the exploration of multi-domain strategies that combine time, power, frequency, and spatial characteristics for RIS identification appears promising. Such an approach could leverage the strengths of each domain while mitigating their individual limitations. 
It should also be noted that the optimal identification domain and technique may vary in single and multi-RIS zones.


\subsection{Other Challenges}





Developing RIS identification techniques that seamlessly operate within current and future wireless standards without extensive modifications is critical. These approaches should be scalable and backward compatible to meet the short-term and long-term needs of various wireless networks in real-time, reducing the integration burden on beyond 5G networks. Key considerations include:


\begin{enumerate}

\item Environmental and Physical Limitations: Factors such as multipath propagation, physical obstructions, and varying atmospheric conditions can affect the reliability and accuracy of RIS identification. Developing robust identification techniques that can adapt to or overcome these environmental challenges is essential for ensuring consistent network performance.


\item Security and Privacy Concerns: It is of
great importance to ensure that the process of identifying RIS panels does not become a vector for malicious activities or privacy breaches. Techniques for secure RIS identification and data transmission must be developed, alongside measures to protect against spoofing attacks, where an adversary might mimic an RIS ID to disrupt network operations or intercept communications.


\item Interference Management and Network Coexistence: Developing strategies for interference mitigation and ensuring coexistence among multiple RISs, while maintaining optimal network performance, requires innovative approaches to signal processing and network management. 


\item Compatibility for Network Automation: These techniques must be capable of dynamically adjusting to ensure optimal performance without necessitating constant manual intervention or recalibration.




\item Minimal Overhead Footprint: Ensuring that RIS identification and operation do not adversely affect the user experience or degrade service quality is a significant challenge. Techniques for RIS identification must be efficient and unobtrusive, minimizing any potential disruption to data transmission or delays in service delivery.

\end{enumerate}
Addressing these challenges requires a multidisciplinary approach, leveraging advancements in signal processing, ML, network theory, and cybersecurity. Collaboration among academia, industry, and standardization bodies will be the key for in developing innovative solutions that overcome these hurdles, paving the way for the successful integration of RIS technology into future wireless networks.

\section{Conclusion}

In this article, we have provided comprehensive discussions and a framework for RIS identification for the first time in the literature. Fundamental principles and advantages in wireless networks and the operational impact and strategic importance of RIS identification have been discussed. In addition, we have discussed the issues and challenges inherent in RIS identification and propose potential strategies and future research avenues for realizing RIS identification in different domains, each with various issues and considerations.

\bibliographystyle{IEEEtran}
\bibliography{IEEEabrv,Ref}





	
\end{document}